\documentclass[11pt]{article}
\usepackage[dvips]{graphicx}
\usepackage{latexsym}
\usepackage{amssymb}

\newcommand{\C}{\mathbb{C}}
\newcommand{\CP}{\mathbb{CP}}

\newcommand{\R}{\mathbb{R}}

\renewcommand{\d}{\mathrm{d}}

\newcommand{\koniec}{\begin{flushright}  $\Box $ \end{flushright}}
\def\be{\begin{equation}}
\def\ee{\end{equation}}

\def\u{\bf{u}}

\def\Sm{\Sigma}

\def\Om{\Omega}
\def\Th{\Theta}

\def\om{\omega}

\def\ov{\overline}

\def\p{\partial}

\def\ov{\overline}

\newcommand{\hook}{{\setlength{\unitlength}{11pt}   
                   \begin{picture}(.833,.8)
                   \put(.15,.08){\line(1,0){.35}}
                   \put(.5,.08){\line(0,1){.5}}
                   \end{picture}}}
\def\a{\alpha}
\def\ll{\lambda}

\def\OO{{\cal O}}

\newtheorem{theo}{Theorem}[section]

\begin{document}
\date{November 13, 2003}
\title{A class of Einstein--Weyl spaces associated to an integrable
system of hydrodynamic type.}
\author{Maciej Dunajski\\
Department of Applied Mathematics and Theoretical Physics, \\
Cambridge University,\\
Wilberforce Road, Cambridge, CB3 OWA, UK}
\maketitle
\noindent
\abstract{HyperCR Einstein--Weyl equations in 2+1 dimensions reduce
to a pair of quasi-linear PDEs of hydrodynamic type. All solutions
to this hydrodynamic system can be in principle constructed from a
twistor correspondence, thus establishing the integrability.
Simple examples of solutions including the hydrodynamic reductions
yield new Einstein--Weyl structures.}
\section{The Equation}
Let us consider a pair of quasi-linear PDEs 
\be
\label{PMA}
u_t+w_y+uw_x-wu_x=0,\qquad u_y+w_x=0,
\ee
for two real functions $u=u(x, y, t), w=w(x, y, t)$.
This system of equation has recently attracted a lot of attention in
the integrable systems literature \cite{Pa03,FK03,FK04,MaSh02}. 
In \cite{DPhil} it arouse in a  different context,
as a symmetry reduction of the heavenly equation.

The system (\ref{PMA})  shares many properties
with two more prominent dispersionless integrable equations: the 
dispersionless Kadomtsev--Petviashvili equation (dKP), and 
the $SU(\infty)$ Toda equation, but it is simpler in some ways. 
\begin{itemize}
\item
Its Lax
representation
\be
\label{Lax11}
[L, M]=0, \qquad \mbox{where}\qquad
L=\p_t-w\p_x-\ll\p_y,\;
M=\p_y+u\p_x-\ll\p_x
\ee
does not contain derivatives with respect to the spectral parameter $\ll$
(the Lax pairs for $SU(\infty)$ Toda, and dKP contain such terms).
\item
Consider a one-form 
\[
e(\ll)=\d x-u \d y+w\d t+\ll(\d y-u\d t)+\ll^2\d t.
\]
The system (\ref{PMA}) is equivalent to the Frobenius integrability
condition
\be
\label{dual}
e(\ll)\wedge\d e(\ll)=0,
\ee
where $\d$ keeps $\ll$ constant. 
This formulation  is dual to the Lax representation (\ref{Lax11}), 
because the distribution spanned by $L$ and $M$ can be defined as
the kernel of $e(\ll)$.  The analogous dual formulations of dKP and $SU(\infty)$
Toda involve distributions defined by two--forms \cite{Kri94, Ta90, DMT00}, 
and are considerably more complicated.
\end{itemize}
One of the aims of this paper is to provide a twistor description of 
(\ref{PMA}) given by the following
\begin{theo}
\label{thMD}
There is a one-to-one correspondence between the equivalence classes 
of solutions to {\em(\ref{PMA})} under point transformations,
and complex surfaces 
(twistor spaces) ${\cal Z}$ such that
\begin{itemize}
\item There exists a holomorphic fibration 
$\pi:{\cal Z}\longrightarrow \CP^1$. 
\item There exists a three--parameter family of holomorphic sections of 
$\pi$ with normal bundle $\OO(2)$ invariant under an anti-holomorphic 
involution $\tau:{\cal Z}\longrightarrow{\cal Z}$ which fixes an
equator of  each section.
\end{itemize}
\end{theo}
The existence of the anti-holomorphic map $\tau$ is required to construct
real solutions to (\ref{PMA}). If one is merely 
interested in complex solutions, then the holomorphic fibration, and
its $\OO(2)$ sections are all one needs. 
Theorem \ref{thMD} provides a parametrisation of local solutions to 
a nonlinear eq. (\ref{PMA}) by a holomorphic data unconstrained by any 
equations. In this sense it resembles the Inverse Scattering
Transform. It remains to be seen whether this Theorem  can be
implemented in practise to construct explicit new solutions to
(\ref{PMA}).

The proof of Theorem \ref{thMD} will be postponed to Subsection 
\ref{proofof_th}.
In the next Section we shall demonstrate that solutions of (\ref{PMA})
can be used to construct Lorentzian Einstein--Weyl (EW) structures in three
dimensions (formulae (\ref{PMAEW})). 
All EW structures which admit a hyperboloid of 
Cauchy--Riemann structures locally arise form solutions to
(\ref{PMA}). We shall give examples of new
Einstein--Weyl spaces which arise in that way. In Subsection \ref{hydrored}
we review the hydrodynamic reductions of (\ref{PMA}), and use them 
to construct another class of Einstein--Weyl spaces.
Finally, in Section \ref{sec_hier} we study a hierarchy of commuting flows
associated to (\ref{PMA}).
\section{The Geometry}
Let $W$ be a  $3$-dimensional manifold with a torsion-free connection $D$,
and a conformal structure $[h]$ of signature $(++-)$ 
which is compatible with $D$ in a sense
that
\[
Dh=\om\otimes h
\]
for some one-form $\om$.
Here $h\in[h]$ is a representative metric in a conformal class. If 
we  change this representative by $h\rightarrow \psi^2 h$, 
then $\om\rightarrow \om +2\d \ln{\psi}$, where $\psi$ is a 
non-vanishing function on $W$. A triple $(W, [h], D)$ is called a Weyl 
structure.
The conformally invariant Einstein--Weyl equations are
\be
\label{EW}
R_{(ab)}=\frac{1}{3}Rh_{ab},\qquad a, b, ... =1, 2, 3.
\ee
Here $R_{(ab)}$ is the symmetrised Ricci tensor of $D$, and $R$ is the
Ricci scalar. 
One can regard $h$ and $\om$ as the unknowns in these equations.
Once they have been  found, the covariant differentiation w.r.t $D$ is given by
\[
D\chi=\nabla \chi-\frac{1}{2}(\chi\otimes \om+(1-m)\om\otimes V-h(\om,\chi)h),
\] 
where $\chi$ is a one--form of conformal weight $m$, and $\nabla$ is 
the Levi--Civita connection of $h$.

It is well known \cite{C43, H82, PT93} 
that the EW equations are equivalent to 
the existence of a two  dimensional family of
surfaces $Z\subset W$ which are 
null with respect to $h$, and  totally geodesic with respect to $D$.
This condition has been used in \cite{DMT00} to construct a Lax representation
for EW equation. The details are as follows:
Let $V_1, V_2, V_3$ be three independent 
vector fields on $W$, and let $e_1, e_2, e_3$ be the dual one-forms.
Assume that 
\[
h=e_2\otimes e_2-2(e_1\otimes e_3+e_3\otimes e_1)
\] 
and some one--form $\om$ give an EW structure.
Let $V(\ll)=V_1-2\ll V_2+\ll^2V_3$ where $\ll\in\CP^1$. 
Then $h(V(\ll), V(\ll))=0$ for 
all $\ll\in\CP^1$ so $V(\ll)$ determines a sphere of null vectors. 
The vectors $V_1-\ll V_2$ and
$V_2-\ll V_3$ form  a basis of the orthogonal complement
of $V(\ll)$. For each $\ll\in\CP^1$ they span 
a null two-surface. 
Therefore  the Frobenius theorem implies that the 
horizontal lifts 
\be
\label{EWlax}
L=V_1-\ll V_2+l\p_{\ll},\qquad
M=V_2-\ll V_3+m\p_{\ll}
\ee
of these vectors to $T(W\times\CP^1)$
span an integrable  distribution, and (\ref{EW}) is equivalent to  
\[
[L, M]=\alpha L+\beta M
\]
for some $\a, \beta$ which are linear in $\ll$.
The functions $l$ and  $m$ 
are third order in $\ll$, because the M{\"o}bius transformations of
$\CP^1$ are generated by  vector fields quadratic in $\ll$.

Let $W_1, ..., W_4$ be  linearly dependent vector fields  
which span $TW$. Given a Lax representation   $[W_1-\ll W_2, W_3-\ll W_4]=0$, 
it is always possible to put it in the form (\ref{EWlax}) with $m=l=0$.
Therefore the   Lax pair (\ref{Lax11}) 
for equation (\ref{PMA}) is a special case of the 
Einstein--Weyl Lax pair (\ref{EWlax}). One finds that
\[
V_1=\p_t+u\p_y+(u^2-w)\p_x, \qquad V_2=\p_y+u\p_x,\qquad V_3=\p_x,
\]
and
\[
[V_1-\ll V_2, V_2-\ll V_3]=-(V_2(u)-\ll V_3(u))(V_2-\ll V_3)
\]
is equivalent to equation (\ref{PMA}).

The dual one--forms $(e_1, e_2, e_2)$ give a metric in the EW
conformal class. The associated one--form can now be found such that
the resulting  Einstein--Weyl structure is
\be
\label{PMAEW}
h=(\d y-u\d t)^2-4(\d x-u \d y+w\d t)\d t,\qquad 
\om =u_x\d y+(uu_x+2u_y)\d t.
\ee
The Ricci scalar of $D$ is 
$
R=(3/8)(u_x)^2,
$
and the one--form $e(\ll)$ used in the dual formulation (\ref{dual}) is 
$e(\ll)=e_1+2\ll e_2+\ll^2 e_3$.

The absence of the vertical terms in the Lax pair implies that
the Einstein--Weyl structure belongs to the Lorentzian analogue of the 
so called hyperCR class \cite{GT98}.  The (Lorentzian) hyperCR EW spaces
arise on the space of trajectories of tri-holomorphic 
conformal Killing vectors in  four-dimensional manifolds 
with (pseudo) hyper--complex structure\footnote{A smooth real $4$-dimensional manifold ${\cal M}$ equipped with three real 
endomorphisms $I, S, T:T{\cal M}\rightarrow T{\cal M}$
of the tangent bundle 
satisfying the algebra of pseudo-quaternions 
\[
-I^2=S^2=T^2=1,\qquad IST=1,
\]
is called
pseudo-hyper-complex iff the almost complex structure $
{\bf J}=aI+bS+cT
$
is integrable for any point of the hyperboloid 
$a^2-b^2-c^2=1$.
A choice of a vector $X\in T{\cal M}$  defines a $(++--)$
conformal structure $[g]$ with an orthonormal frame
$X, IX, SX, TX$. If this conformal structure admits a Killing vector
$K$ which preserves $I, S, T$, then the hyperboloid of complex
structures ${\bf J}$ descends to a hyperboloid of 
Cauchy--Riemann (CR) structures on the space of orbits $W$ of $K$.
This justifies the terminology.}.
Readers unfamiliar with the details of these pseudo--hyper--complex
geometries in four--dimensions should note that 
the condition $l=m=0$ in (\ref{EWlax})  can be used as 
an equivalent  definition of the hyperCR class, and can go directly 
to the statement   of Theorem \ref{theoEWCR}.

Any pseudo--hyper--complex conformal structure $([g], I, S, T)$ 
in four dimensions 
with a conformal
Killing vector $K$ gives rise to an EW structure \cite{JT85,DMT00}
defined by
\be
\label{EWs}
h:=|K|^{-2}g-|K|^{-4}{K}\otimes {K},\qquad  \om:=2|K|^{-2}
\ast_g({K}\wedge \d{K}), \qquad g\in[g],
\ee
and all EW spaces arise form this construction.
If the Killing vector $K$ preserves the endomorphisms
$I, S, T$, then the resulting EW structure (\ref{EWs}) is called
hyperCR. 
Conversely, given a hyperCR Einstein--Weyl structure $(h, \om)$
one can construct the representative $g$ of a pseudo--hyper--complex
conformal class $[g]$ by
\be
\label{metric}
g=e^{T}(Vh-V^{-1}(\d T +\beta)^2),
\ee
where $T$ is a group parameter, and 
the function $V$ and the one--form $\beta$  solve  
the monopole equation
\be
\label{monopoleCR}
*(\d V+(1/2)\om V) =\d\beta
\ee
(here $*$ is taken with respect to $h$). 
There exists a special solution to (\ref{monopoleCR}) with $\beta=-\om/2$,
such that the resulting metric is pseudo--hyper--Kahler (the two--forms
associated to $I, S, T$ are closed).  The details 
of all that are in  \cite{GT98}. Minor sign changes are 
needed to apply the theory in signature $(++--)$. The hyperCR EW
structure were constructed out of symmetry reductions of heavenly
equations in \cite{DT01}.

We shall now show that the system (\ref{PMA}) arises as a symmetry reduction of
the pseudo hyper--K\"ahler condition with by a general 
homothetic Killing vector, thus establishing the following result
\begin{theo}
\label{theoEWCR}
All Lorentzian hyperCR Einstein--Weyl structures are locally of the form 
{\em(\ref{PMAEW})}, 
where $u, w$ satisfy {\em(\ref{PMA})}.
\end{theo}
{\bf Proof.}
It follows form the
work of Pleba\'nski \cite{Pl75} that all 
pseudo hyper--K\"ahler (or ASD vacuum)  metrics are locally of the form
\be
\label{Plebanm}
g=2(\d Z\d Y +\d W\d X -\Th_{XX}\d X^2-\Th_{YY}\d W^2+2\Th_{XY}\d W\d Z),
\ee
where  $(W, Z, X, Y)$ are the local coordinates on the open ball in 
$\R^4$, and  $\Th=\Th(W, Z, X, Y)$
satisfies  the second heavenly equation 
\be
\label{secondeq}
\Th_{ZY}+\Th_{WX}+\Th_{XX}\Th_{YY}-\Th_{XY}^2=0.
\ee
In an ASD vacuum the most general tri-holomorphic homothetic 
Killing  vector $K$ satisfies ${\cal L}_K\Sm_i=c \Sm_i$, where 
${\cal L}_K$ is the Lie derivative along $K$ and  $\Sm_1, \Sm_2,
\Sm_3$ are three closed self-dual two forms corresponding
to the complex structures. We can set $c=1$ without  loose
of generality. In the coordinate system adopted to (\ref{Plebanm})
\[
\Sm_1=\d W\wedge \d Z,\qquad \Sm_2=\d W\wedge \d X +\d Z\wedge \d Y,
\]
and the residual freedom in the choice of coordinates can be used to set
\[
K=Z\frac{\p}{\p Z}+X\frac{\p}{\p X}.
\]
The Killing equations yield
\[
{\cal L}_K(\Th_{XX})=-\Th_{XX}, \qquad
{\cal L}_K(\Th_{XY})=0, \qquad
{\cal L}_K(\Th_{YY})=\Th_{YY}.
\]
Let $U$ and $T$ be functions on $\R^4$ such that $K=\p/\p T$ and
${\cal L}_K(U)=0$. We can take
\[
T=\ln{(Z)},\qquad U=-\frac{X}{Z}.
\]
The compatibility conditions for the Killing equations imply the
existence of ${G}={G}(Y, W, U)$ such that
\[
\Th_{XX}=-e^{-T}{G}_{UU}, \qquad
\Th_{XY}={G}_{YU}, \qquad \Th_{YY}=-e^{T}{G}_{YY}.
\]
The heavenly equation (\ref{secondeq}) becomes
\[
-({G}_Y-U{G}_{YU})+
{G}_{UW}+{G}_{YY}{G}_{UU}-{G}_{YU}^2=0,
\]
or (in terms of differential forms)
\be
\label{form2}
-G_Y\d Y\wedge\d U\wedge\d W+U\d G_U\wedge\d U\wedge\d W
+\d G_U\wedge \d Y\wedge\d U+\d G_Y\wedge\d G_U\wedge\d W.
=0
\ee
Define 
\[
x=G_U,\qquad y=Y
,\qquad t=-W,\qquad H(x, y, t)=xU(x, y, t) 
-G(Y, W, U(x, y, t)), 
\]
and preform a Legendre transform
\begin{eqnarray*}
\d H&=&\d (xU-G)=U\d x- G_Y\d Y- G_W\d W\\
&=&H_x\d x+ H_y\d y+H_t\d t.
\end{eqnarray*}
Therefore
\[
U=H_x, \qquad G_Y=-H_y, \qquad G_W=H_t.
\]
Differentiating these relations we find
\[
G_{UU}=\frac{1}{H_{xx}},\qquad
G_{YU}=-\frac{H_{xy}}{H_{xx}},\qquad
G_{YY}=-H_{yy}+\frac{H_{xy}^2}{H_{xx}}.
\]
The differential equation for $H(x, y, t)$ is obtained form (\ref{form2})
\be
\label{Heq1}
H_{xt}-(H_{xy}H_x-H_yH_{xx})=H_{yy}.
\ee
This equation is equivalent to the system (\ref{PMA}) which can be seen by
setting $u=H_x, w=-H_y$.

The metric (\ref{Plebanm}) can be written in the form
(\ref{metric})
where $h, \om$ are given by (\ref{PMAEW}), and $V=u_x/2, \beta=-\om/2$ 
satisfy   the monopole equation (\ref{monopoleCR}).
We deduce that (\ref{PMAEW})
is the most general EW space which arises on the space of orbits of
tri-holomorphic homothety in pseudo--hyper--K\"ahler four manifold,
and so it is the most general hyperCR EW space.
\koniec
\subsection{Simple solutions}
Simple classes of solutions to (\ref{PMA}) yield nontrivial
Einstein--Weyl structures, some of which appear to be new.
\begin{itemize}
\item
Let us assume that $u$ and $w$ in (\ref{PMA}) do not depend on $y$. 
One needs to consider the two cases $w=0$ and $w=w(t)\neq 0$ separately.
The corresponding equations can now be easily integrated to give 
(in the $w\neq 0$ case one needs to change variables)
\be
\label{newEW}
h=(\d y+A\d t)^2-4\d x\d t, \qquad \om=A'\d y+AA'\d t,
\ee
where $A=A(x)$ is an arbitrary function. Some interesting 
complete solutions belong to this class. For example 
$A=a^2x$, where $a$ is a non-zero constant leads
to the Einstein--Weyl structure  on Thurston's Nil manifold 
$S^1\times \R^2$ \cite{PT93}: Setting $\hat{x}=a^2x$, and rescalling $h$ by a 
constant factor gives 
\[
h=a^2(\d y+\hat{x}\d t)^2-4\d \hat{x}\d t,\qquad 
\om=a^2(\d y+\hat{x}\d t).
\]
In this simple case we can find a kernel of the Lax vector fields 
(\ref{Lax11}) (the twistor functions) to be 
$\ll, \psi=y+\ll t-a^{-2}\ln{(\ll-a^2x)}$.
\item
Looking for $t$--independent solutions reduces (\ref{PMA}) to a
linear equation. Rewriting the resulting system of as
\[
\d x\wedge\d u-\d y\wedge\d w=0,\qquad \d x\wedge\d w-(u\d w-w\d
u)\wedge\d y=0,
\]
and regarding $x$ and $y$ as functions of $u$ and $w$ yields a system
of linear equation. One of these equations implies that $y=-F_w,
x=F_u$ for some $F=F(u,w)$, while the other equation yields
\[
F_{uu}+uF_{uw}+wF_{ww}=0.
\]
\item
The constraint $u_x=0$ leads to trivial EW spaces. One finds
that $u$ has to be linear in $y$, and the EW one--form $\om$ is closed.
It can therefore be set to $0$ be the conformal rescaling, and the
EW structure is conformal to an Einstein metric.
\end{itemize}
\subsection{Proof of Theorem (\ref{thMD})}
\label{proofof_th}
Given a real--analytic solution to (\ref{PMA}) we can complexify it, and
and regard $u$ and $w$ as 
holomorphic functions of local 
complex coordinates $(x, y, t)$ on a complex
three--manifold $W^{\C}$. The twistor space ${\cal Z}$ for such solution 
is obtained by
factoring $F=W^{\C}\times \CP^1$ by the 
distribution $L, M$ (\ref{Lax11}). This clearly has a projection
$q :F \mapsto {\cal Z}$ and we have a double
fibration
$$
\begin{array}{rcccl}
&&W^{\C}\times\CP^1&&\\
&r\swarrow&&\searrow q&\\
&W^{\C}&&{\cal Z}.&
\end{array}
$$
The absence of vertical terms in $L, M$ shows that $\ll$ descends from
$F$ to ${\cal Z}$ thus giving the holomorphic projection 
$\pi:{\cal Z}\longrightarrow \CP^1$. Each 
point $p\in W^{\C}$ determines a sphere $l_p$ (a section of $\pi$) 
made up of all the
integral surfaces of $L, M$ through $p$. 
The normal bundle of $l_p$
in $\cal Z$ is $N=T{\cal Z}|_{l_p}/Tl_p$. This is a rank one vector
bundle over $\CP^1$, therefore it has to be one of the standard line
bundles ${\cal O}(n)$. To see that $n=2$, 
note that $N$ can be identified with the
quotient $r^*(T_p{W^{\C}})/\{ \mathrm{span }\;L,M\}$.
In their homogeneous form the operators $L, M$ have weight one, so the
distribution spanned by them is isomorphic to the bundle
$\C^{2}\otimes{\cal O}(-1)$.  The definition of the normal bundle as
a quotient gives a sequence of sheaves over $\CP^1$.
\[
0\longrightarrow \C^{2}\otimes{\cal O}(-1) \longrightarrow \C^{3}  
\longrightarrow N\longrightarrow 0
\]
and we see that $N={\cal O}(2)$, because the
last map,  
is given explicitly by $(V_1, V_2, V_3)\mapsto V(\ll)=V_1-2\ll V_2+\ll^2V_3$ 
clearly projecting onto ${\cal O}(2)$.

If $u, w$ is a real solutions defined on a real slice $W\subset W^{\C}$,
then one has an additional structure on ${\cal Z}$:
The real structure $\tau(x, y, t)=(\overline{x}, \ov{y}, \ov{t})$ 
maps integral surfaces of $L, M$ to integral surfaces, 
and therefore induces an anti-holomorphic
involution $\tau:  {\cal Z}\rightarrow {\cal Z}$.
The fixed points of this involution  correspond to real  
integral surfaces in $W$, and $\tau$--invariant $\OO(2)$ sections
correspond to points in $W$.

 Conversely, let us assume that we are given a complex manifold ${\cal Z}$
with additional structures described in Theorem \ref{thMD}. The general 
construction of Hitchin \cite{H82} equips the moduli 
space $W$ of $\OO(2)$ rational curves with a real-analytic 
Einstein--Weyl structure: The Kodaira theorems imply that $W$ is 
three dimensional. Two points in $W$ are null-separated if the corresponding
sections intersect at one point in ${\cal Z}$. This defines
the conformal structure $[h]$. To define a connection note that a direction
at $p\in W$ corresponds to a one--dimensional space of $\OO(2)$ curves in 
${\cal Z}$ which vanish at two points $Z_1$ and $Z_2$. This gives distinguished
curves in $W$ which pass through null surfaces in $W$ 
corresponding to $Z_1, Z_2$.
There is one such curve through $p$ and Hitchin defines it to be a geodesic.
He moreover shows that the resulting connection is torsion--free, and that
the Einstein--Weyl equations hold. 

This works for arbitrary complex surface with an embedded $\OO(2)$ rational 
curve. The additional structure in the statement of Theorem \ref{thMD} is
the holomorphic projection $\pi$. Its existence implies that the resulting
EW space is hyper--CR.
Any  holomorphic line bundle
$L\longrightarrow\OO(2)$ with $c_1(L)=0$ inherits the holomorphic projection
to $\CP^1$. Lifts of holomorphic sections of 
${\cal Z}\longrightarrow \CP^1$ to $L$ are rational curves with normal bundle $\OO(1)\oplus\OO(1)$. Therefore $L$ is 
a twistor space of a pseudo-hyper-complex  four-manifold ${\cal M}$ \cite{D99}.
This pseudo hyper--complex structure is preserved by a Killing vector
which gives rise to  a hyperboloid of CR structures on $W$.
The Theorem \ref{theoEWCR} implies that $(h, \om)$ are locally given by 
(\ref{PMAEW}), and we can read off $(u, w)$ which solve equation (\ref{PMA}).
\koniec
\subsection{Geodesic Congruences}
Let $(W, [h], D)$ be a 2+1 EW structure.
A geodesic congruence $\Gamma$ in a region
in $\hat{W}\subset W$ is a set of geodesic, one through each point
of ${\hat W}$. Let $\chi$ be a generator of $\Gamma$ (a vector field tangent
to $\Gamma$). The geodesic condition $\chi^aD_a\chi^b\sim \chi^b$ implies 
$D_a\chi^b={M_a}^b+A_a\chi^b$ for some $A_a$, where ${M_a}^b$ is orthogonal to 
$\chi^a$ on both indices. Consider the decomposition of $M_{ab}$
\[
M_{ab}=\Om_{ab}+\Sm_{ab}+\frac{1}{2}\theta \hat{h}_{ab}.
\]
The shear $\Sm_{ab}$  is trace-free and symmetric.
The twist $\Om_{ab}$ is anti-symmetric, and the divergence $\theta$ is 
is a weighted scalar.
Here $\hat{h}_{ab}=||\chi||^2h_{ab}-\chi_a\chi_b$ is 
an orthogonal projection of $h_{ab}$.
The 
shear-free geodesics congruences (SFC) exist on any Einstein--Weyl space.
This follows from a three-dimensional version of Kerr's theorem which 
states that SFCs correspond to  
to holomorphic curves in the twistor space ${\cal Z}$.
On the other hand imposing the vanishing of the divergence of a congruence 
gives restrictions on EW structures, and implies that the EW space
is hyperCR \cite{CP}. In the local coordinate system adopted in
Theorem \ref{theoEWCR} $(h, \om)$ are given by (\ref{PMAEW}), and 
the  shear-free, divergence free geodesic congruence is generated by a
one form $\chi=\d t$.  In accordance with the general theory of 
SFC on Einstein--Weyl spaces \cite{CP},
the preferred monopole proportional to the 
scalar twist $\kappa=*(\chi\wedge D\chi)=-u_x/4$ will lead to a
pseudo--hyper--K\"ahler metric with  a tri--holomorphic 
homothety in four dimensions. This metric is explicitly given by 
(\ref{metric}), where $V=-\frac{1}{2}\kappa, \beta=-\om/2$. 
Any other monopole yields a general 
pseudo--hyper--complex conformal structure with a tri--holomorphic symmetry.
\section{The hydrodynamic reductions}
\label{hydrored}       
The equation (\ref{PMA}) can be cast in a general quasi-linear vector form
\be
\label{qle}
{\u}_y+A({\u}){\u}_x+B({\u}){\u}_t=0,
\ee
where ${\u}=(u, w)^T$ is a vector whose
components depend on $(x, y, t)$,  
and  
\[
A({\u})=
\left (
\begin{array}{cc}
0&1\\
-w&u
\end{array}
\right ),\qquad  B({\u})=
\left (
\begin{array}{cc}
0&0\\
1&0
\end{array}
\right ).
\] 
A class of solutions to any equation of the form (\ref{qle}) 
can be generated by assuming that  
${\u}={\u}(R^1, ... ,R^N)$, where  $R^i=R^i(x, y, t)$ 
(the so called Riemann invariants)
satisfy a pair of commuting systems of hydrodynamic type
\be
\label{Riem}
{R^i}_y=\gamma^i(R){R^i}_x, \qquad {R^i}_t=\mu^i(R){R^i}_x,\qquad
i=1, 2, ..., N
\ee
(the summation convention has been suspended in this section).
The compatibility conditions for the system (\ref{Riem}) yield
\be
\label{compatibility}
\frac{\p_j\gamma^i}{\gamma^j-\gamma^i}=
\frac{\p_j\mu^i}{\mu^j-\mu^i}, \qquad i\neq
j,\qquad
\p_j=\frac{\p}{\p R^j}.
\ee
It turns out that the additional relations 
\[
\p_k\frac{\p_j\gamma^i}{\gamma^j-\gamma^i}=\p_j\frac{\p_k\gamma^i}{\gamma^k-
\gamma^i},
\]
(and analogous relations for $\mu^i$) hold. 
These conditions imply th existence of a diagonal metric
$g=\sum_ig_{ii}(R)\d (R^i)^2$  such that $\Gamma_{ji}^i
=\p_j\ln{(\sqrt{g_{ii}})}=\p_j\gamma^i/(\gamma^j-\gamma^i)$ are the contracted
components of the Levi--Civita connection of $g$. 

If conditions (\ref{compatibility}) are satisfied, 
the general solutions to (\ref{Riem})
is implicitly given by the generalised hodograph formula
of Tsarev \cite{Ts90}
\[
v^i(R)=x+\gamma^i(R)y+\mu^i(R)t, \qquad i=1, ..., N.
\]
Once $\gamma^i$ have been found,
the functions $v^i$ (called the characteristic speeds) should be determined 
from the linear relations  $\Gamma_{ji}^i=\p_j v^i/( v^j- v^i), i\neq j.$
Substituting this expression in (\ref{qle}) shows that $\p_i\u$ are 
eigenvectors of $(A-\gamma I-\mu B)$ with zero eigenvalue. 
Therefore $\gamma^i$ and $\mu^i$ satisfy the dispersion relation
\be
\label{dispersion}
\det(A-\gamma I-\mu B)=0.
\ee
Solutions to (\ref{qle}) obtained form this algorithm are known as the
non--linear interactions of $N$ planar simple waves.
The procedure explained in this sections has been applied 
in \cite{GibTsa96, Fer, Fer2} to construct explicit solutions to
various PDEs which admit a representation (\ref{qle}).

Ferapontov and Khusnutdinova \cite{FK03} define a hyperbolic system 
of the form (\ref{qle}) to be
integrable if it possesses non-linear interactions of
$N$ planar simple waves parametrised by $N$ arbitrary functions of one 
variable.
They have 
demonstrated \cite{FK04} that this definition of integrability is
equivalent to the existence a scalar pseudo-potential formulation  of
the form
\[
\psi_y=P(\psi_x, u, w), \qquad \psi_t=Q(\psi_x, u, w),
\]
where $\psi=\psi(x, y, t)$ and $P, Q$ are rational in $\psi_x$.
This then implies that the integrable equations (\ref{qle})
arise as dispersionless (or quasiclassical) limits of non-linear
PDEs solvable by Inverse Scattering Transform \cite{Z94}.
These `dispersive' PDEs are compatibility conditions for the
overdetermined linear system
\[
\Psi_Y=P(\p/\p X)\Psi, \qquad \Psi_T=Q(\p/\p X)\Psi,
\]
where now $P$ and $Q$ are linear differential operators, and the 
dispersionless limits can be obtained by setting
\[
\frac{\p}{\p X^a}\longrightarrow\varepsilon\frac{\p}{\p x^a}, \qquad
\Psi(X^a)=\exp{(\psi(x^a/\varepsilon)},
\]
and taking the limit $\varepsilon\longrightarrow 0$.
Finding a dispersive analogue of (\ref{PMA}) is an interesting open problem.

\subsection{Example}
According to Pavlov \cite{Pa03} the hydrodynamic reductions of (\ref{PMA})
are characterised in a sense that
explicit formulae for $\gamma^i(R)$, and
$\mu^i(R)$ can be found. 
This does not however lead to explicit (or even implicit) solutions to
(\ref{PMA}). The constraints on a solution to (\ref{PMA}) 
imposed by the existence of $N$--component reductions are not known. 
To this end, we shall work out the constraint, and the 
corresponding solution which arise form 
a one-component reduction. 

For $N=1$ we have $u=u(R), w=w(R)$ where the
scalar variable $R=R^1$ satisfies a pair of PDEs 
$R_y=\gamma(R)R_x, R_t=\mu(R)R_x$. All integrability conditions hold automatically, and the dispersion relation 
(\ref{dispersion}) yields 
\[
\mu=w+\gamma u+\gamma^2.
\]
Implicit differentiation of $u, w$ with respect to $(x, y, t)$, and 
eliminating $(u', w', R_x)$  gives
a constraint 
\be
\label{const1}
u_xw_y-u_yw_x=0
\ee
which characterises solutions to (\ref{PMA})  arising form
one-component hydrodynamic reductions. 
Using the relations
\[
*\d t=\d t \wedge \d y,\qquad *\d y=2\d t \wedge \d x-u\d t\wedge \d y,\qquad
*\d x=2w\d y\wedge\d t+\d y\wedge\d x+u\d t\wedge\d x,
\]
where $*:\Lambda^a(W)\longrightarrow \Lambda^{3-a}(W)$ is the Hodge operator
associated to the EW structure (\ref{PMAEW}) we find that the constraint
(\ref{const1}) is equivalent to the relation
\[
|\d u|^2:=\d u\wedge *\d u=0. 
\]
The solution can now  easily be found
by applying the Legendre transform. Regarding $u$ and $y$ as functions of
$(w, t, x)$ gives
\[
u_x=0, \qquad  u_w=y_x, \qquad
u_wy_t-u_ty_w-1+uy_x-w(u_wy_x-u_xy_w)=0,
\]
where the first relation arises form the constraint (\ref{const1}).
These equations can be integrated to give two classes of solutions 
\[
u_1=at+aw+\frac{1}{a},\; y_1=ax-\frac{at^2}{2}+f_1(w+t),\qquad
u_2=2\sqrt{w+t}, \; y_2=\frac{x-t}{\sqrt{w+t}}+f_2(w+t).
\]
where $f_1$ and $f_2$ are  arbitrary functions of one variable
(one arbitrary function of $t$ has been eliminated in each class 
by a redefinition 
of coordinates), and $a$ is a non-zero constant. 
The resulting Einstein--Weyl spaces (\ref{PMAEW})
can be written down explicitly, and are 
completely characterised by the condition $|\d u|=0$.
\section{The Hierarchy}
\label{sec_hier}
Consider a sphere of one--forms on an open set in $\R^{n+1}$
\[
e(\ll)=\d t_0+(\ll-H_0)\d t_1+(\ll^2-\ll H_0-H_1)\d
t_2+...+(\ll^n-\ll^{n-1}H_0-...-\ll H_{n-2}-H_{n-1})\d t_n,
\]
where $H=H(t_0, t_1, ..., t_n), H_a=\p H/\p t_a$ and $\ll\in \CP^1$.
The system of PDEs
\be
\label{hierarchy}
e(\ll)\wedge\d(e(\ll))=0
\ee
coincides with (\ref{PMA}) if $n=2, t_0=x, t_1=y, t_2=t$ and
$u=H_x, w=-H_y$. If $n>2$ then (\ref{hierarchy}) is highly
overdetermined, and the Cauchy data can be specified on a
surface of co--dimension $n-1$ (rather than on a hypersurface).
We shall call this system a truncated hierarchy associated to
(\ref{PMA}). Allowing infinite sums in $e(\ll)$ would lead to
the full hierarchy. The Frobenius theorem implies that
an $n$--dimensional distribution of vector fields on 
$\R^{n+1}\times \CP^1$ annihilating $e(\ll)$ is in involution. This
gives rise to the Lax representation. 
The vector fields
\be
\label{lax_hier}
L_a=\frac{\p}{\p t_{a+1}}+\frac{\p H}{\p t_a}\frac{\p}{\p t_0}-\ll
\frac{\p}{\p t_a}, \qquad a=0, ..., n-1
\ee
satisfy $L_a\hook e(\ll)=0$, and the relations
\[
[L_a, L_b]=0
\]
yield the commuting flows of the hierarchy
\be
\label{hierarchy_ex}
H_{(a+1)b}-H_{(b+1)a}+H_aH_{0b}-H_bH_{0a}=0.
\ee 
The Lax representation (\ref{lax_hier}) fits into a general class
of Lax formulations recently introduced in \cite{MaSh02}.
The Theorem \ref{thMD} should generalise to solutions of
(\ref{hierarchy_ex}). The twistor space  ${\cal Z}$ 
is  a surface  which  arises as a factor space
of $\C^n\otimes\CP^1$ by a complexified distribution (\ref{lax_hier}).
Repeating the steps of the proof of Theorem \ref{thMD} shows that
the holomorphic fibration ${\cal Z}\longrightarrow \CP^1$ 
admits an $(n+1)$ family
of holomorphic sections with normal bundle $\OO(n)$. 
The converse (recovering $H(t_0, t_1, ..., t_n)$) form ${\cal Z}$ is
however more difficult, because the vital relation (\ref{PMAEW}) 
with Einstein--Weyl
geometry is missing for $n>2$. This interesting problem, and
its connection with the quasiclassical $\ov{\p}$ approach \cite{Kon3} will be
addressed elsewhere.

\section*{Acknowledgements}
I thank Jenya Ferapontov for useful discussions about the 
hydrodynamic reductions. This research was partly 
supported by NATO grant PST.CLG.978984, and CONACYT grant 41993-F.

\end{document}